\documentclass[12 pt,twoside,aps,prd,amsmath,amssymb,
doublespace,showpacs,showkeys,eqsecnum]{revtex4-1}

\usepackage{xcolor}

\usepackage{graphicx}
\usepackage{mathptmx}
\usepackage{bm}
\newcommand {\pr}{\partial}
\newcommand {\cG}{\cal G}
\newcommand {\cD}{\cal D}
\newcommand {\bg}{\bar \gamma}
\newcommand {\G}{\Gamma}
\newcommand {\bp}{\bar \psi}
\newcommand {\p}{\psi}

\def\myfigure #1#2#3#4
{\begin{figure}[ht]\begin{center}
\includegraphics[width=#2 \textwidth]{#1.eps}
\parbox[t]{#4\textwidth}{\caption{#3}\label{#1}}
\end{center}\end{figure}}

\def \myfigures #1#2#3#4#5#6#7#8
{\begin{figure}[ht]
    \begin{center}
        \includegraphics[width=#2 \textwidth]{#1.eps}
        \hfill
        \includegraphics[width=#6 \textwidth]{#5.eps}
        \parbox[t]{#4\textwidth}{\caption {#3}\label{#1}}
        \hfill
        \parbox[t]{#8\textwidth}{\caption {#7}\label{#5}}
    \end{center}
\end{figure} }

\begin{document}
\title{Spinor field with polynomial
nonlinearity in LRS Bianchi type-I space-time}
\author{Bijan Saha}
\affiliation{Laboratory of Information Technologies\\
Joint Institute for Nuclear Research\\
141980 Dubna, Moscow region, Russia} \email{bijan@jinr.ru}
\homepage{http://spinor.bijansaha.ru}

\begin{abstract}

Within the scope of the locally rotationally symmetric (LRS) Bianchi
type-I cosmological model the role of spinor field on the evolution
of the Universe is investigated. In doing so, we have considered a
polynomial type of nonlinearity. It is found that, depending on the
sign of the self-coupling constant, the model allows either an
accelerated mode of expansion or an oscillatory mode of evolution.
While the non-diagonal components of the energy-momentum tensor of
the spinor field in the case of a full Bianchi type-I  model lead to
the vanishing mass and nonlinear term in the spinor field
Lagrangian, in the case of an LRS Bianchi type-I  model neither the
mass term nor the nonlinear term of the spinor field vanish.

\end{abstract}



\maketitle

\bigskip

\section{Introduction}

Recently, after some remarkable works  by different authors
\cite{henneaux,ochs,saha1997a,saha1997b,saha2001a,greene,saha2004a,
saha2004b,ribas,saha2006c,saha2006e,saha2007,saha2006d,souza,kremer},
showing the important role that spinor fields play on the evolution
of the Universe, the situation began to change. This change of
attitude is directly related to some fundamental questions of modern
cosmology: (i) problem of initial singularity; (ii) problem of
isotropization and (iii) late time acceleration of the Universe.

Given the role that the spinor field can play in the evolution of
the Universe, the question that naturally emerges is, if the spinor
field can redraw the picture of evolution caused by perfect fluid
and dark energy, is it possible to simulate perfect fluid and dark
energy by means of a spinor field? An affirmative answer to this
question was given in a number of papers
\cite{krechet,saha2010a,saha2010b,saha2011,saha2012}. In those
papers, a spinor description of matter, such as a perfect fluid and
dark energy, was given and the evolution of the Universe, given by
different Bianchi models, was thoroughly studied. In almost all the
papers the spinor field was considered to be a time-dependent
function and its energy-momentum tensor was given by the diagonal
elements only.

Some latest studies show that because of the specific connection
with the gravitational field the energy-momentum tensor of the
spinor field possesses non-trivial non-diagonal components as well,
and these non-zero non-diagonal components of the energy-momentum
tensor play decisive a role in the character of the geometry of
space-time as well as on the components of the spinor field
\cite{sahaIJTP2014,sahaAPSS2015,sahabvi0,sahabvi}.

In this paper we study the evolution of the Universe filled with
spinor field within the scope of a locally rotationally symmetric
(LRS) Bianchi type-I (BI) cosmological model. It should be noted
that a general BI model in the presence of a nonlinear spinor field
duly evolves into a LRS BI model \cite{sahaAPSS2015}. In this paper
we thoroughly study the role of a spinor field in the evolution of
the Universe given by a LRS BI model. Here we also consider a more
general type of nonlinearity.

\section{Basic equations}

In this paper we plan to study the evolution of the Universe given
by a LRS BI anisotropic cosmological model filled with nonlinear
spinor field.

The LRS BI model is the ordinary BI model with two of the three
metric functions being equal to each other and can be given by
\begin{equation}
ds^2 = dt^2 - a_1^{2} \left[dx^{2}\, + \,dy^{2}\right] -
a_3^{2}\,dz^2, \label{lrsbi}
\end{equation}
with $a_1$ and $a_3$ being functions of time only.

The nontrivial components of the Einstein tensor corresponding to
metric (\ref{lrsbi}) are

\begin{subequations}
\label{LRSBIET}
\begin{eqnarray}
G_1^1 &=& G_2^2 = - \left( \frac{\ddot a_3}{a_3} +\frac{\ddot
a_1}{a_1} + \frac{\dot
a_3}{a_3}\frac{\dot a_1}{a_1}\right),\label{LRSBIET11}\\
G_3^3 &=& - \left(2\frac{\ddot a_1}{a_1} + \frac{\dot
a_1^2}{a_1^2}\right),\label{LRSBIET33}\\
G_0^0 &=& - \left(\frac{\dot a_1^2}{a_1^2} + 2\frac{\dot
a_3}{a_3}\frac{\dot a_1}{a_1}\right). \label{LRSBIET00}
\end{eqnarray}
\end{subequations}

Keeping in mind the symmetry between $\p$ and $\bp$ we choose the
symmetrized Lagrangian \cite{kibble} for the spinor field as
\cite{saha2001a}:
\begin{equation}
L = \frac{\imath}{2} \biggl[\bp \gamma^{\mu} \nabla_{\mu} \psi-
\nabla_{\mu} \bar \psi \gamma^{\mu} \psi \biggr] - m_{\rm sp} \bp
\psi - F, \label{lspin}
\end{equation}
where the nonlinear term $F$ describes the self-interaction of a
spinor field and can be presented as some arbitrary function of
invariants generated from the real bilinear forms of a spinor field.
We consider $F = F(K)$, with $K$ taking one of the following
expressions $\{I,\,J,\,I+J,\,I-J\}$. It can be shown that such a
choice describes the nonlinearity in its most general form.

Here $\nabla_\mu$ is the covariant derivative of spinor field
\begin{equation}
\nabla_\mu \psi = \frac{\partial \psi}{\partial x^\mu} -\G_\mu \psi,
\quad \nabla_\mu \bp = \frac{\partial \bp}{\partial x^\mu} + \bp
\G_\mu, \label{covder}
\end{equation}
with $\G_\mu$ being the spinor affine connection. In (\ref{lspin})
$\gamma$'s are the Dirac matrices in curve space-time and obey the
following algebra:
\begin{equation}
\gamma^\mu \gamma^\nu + \gamma^\nu \gamma^\mu = 2 g^{\mu\nu}
\label{al}
\end{equation}
and are connected with the flat space-time Dirac matrices $\bg$ in
the following way
\begin{equation}
 g_{\mu \nu} (x)= e_{\mu}^{a}(x) e_{\nu}^{b}(x) \eta_{ab},
\quad \gamma_\mu(x)= e_{\mu}^{a}(x) \bg_a, \label{dg}
\end{equation}
where $\eta_{ab}= {\rm diag}-(1,-1,-1,-1)$ and $e_{\mu}^{a}$ is a
set of tetrad 4-vectors. The spinor affine connection matrices
$\G_\mu (x)$ are uniquely determined up to an additive multiple of
the unit matrix by
\begin{equation}
\nabla_\mu \gamma_\nu = \frac{\pr \gamma_\nu}{\pr x^\mu} -
\G_{\nu\mu}^{\rho}\gamma_\rho - \G_\mu \gamma_\nu + \gamma_\nu
\G_\mu = 0, \label{afsp}
\end{equation}
with the solution
\begin{equation}
\Gamma_\mu = \frac{1}{4} \bg_{a} \gamma^\nu \partial_\mu e^{(a)}_\nu
- \frac{1}{4} \gamma_\rho \gamma^\nu \Gamma^{\rho}_{\mu\nu},
\label{sfc}
\end{equation}

The spin affine connection corresponding to LRS BI metric
(\ref{lrsbi}) can be written explicitly as
\begin{equation}
\G_0 = 0, \quad \G_1 = \frac{\dot a_1}{2} \bg^1 \bg^0, \quad  \G_2 =
\frac{\dot a_1}{2} \bg^2 \bg^0, \quad  \G_3 = \frac{\dot a_3}{2}
\bg^3 \bg^0, \label{saclrsbi}
\end{equation}

Varying (\ref{lspin}) with respect to $\bp (\psi)$ one finds the
spinor field equations:
\begin{subequations}
\label{speq}
\begin{eqnarray}
\imath\gamma^\mu \nabla_\mu \psi - m_{\rm sp} \psi - 2 F_K (S K_I +
 \imath P K_J \gamma^5) \psi &=&0, \label{speq1} \\
\imath \nabla_\mu \bp \gamma^\mu +  m_{\rm sp} \bp + 2 F_K \bp(S K_I
+  \imath P K_J \gamma^5) &=& 0. \label{speq2}
\end{eqnarray}
\end{subequations}
Here we denote $F_K = dF/dK$, $K_I = dK/dI$, and $K_J = dK/dJ.$

The energy-momentum tensor of the spinor field is given by
\begin{equation}
T_{\mu}^{\rho}=\frac{i}{4} g^{\rho\nu} \biggl(\bp \gamma_\mu
\nabla_\nu \psi + \bp \gamma_\nu \nabla_\mu \psi - \nabla_\mu \bar
\psi \gamma_\nu \psi - \nabla_\nu \bp \gamma_\mu \psi \biggr) \,-
\delta_{\mu}^{\rho} L_{\rm sp} \label{temsp}
\end{equation}
where $L_{\rm sp}$ in view of (\ref{speq1}) and (\ref{speq2}) can be
rewritten as
\begin{eqnarray}
L_{\rm sp} & = & \frac{\imath}{2} \bigl[\bp \gamma^{\mu}
\nabla_{\mu} \psi- \nabla_{\mu} \bar \psi \gamma^{\mu} \psi \bigr] -
m_{\rm sp} \bp \psi - F(K)
\nonumber \\
& = & \frac{\imath}{2} \bp [\gamma^{\mu} \nabla_{\mu} \psi - m_{\rm
sp} \psi] - \frac{\imath}{2}[\nabla_{\mu} \bar \psi \gamma^{\mu} +
m_{\rm sp} \bp] \psi
- F(K),\nonumber \\
& = & 2 F_K (I K_I + J K_J) - F = 2 K F_K - F(K). \label{lspin01}
\end{eqnarray}

Further inserting (\ref{covder}) into (\ref{temsp}) the
energy-momentum tensor of the spinor field can be written as
\begin{eqnarray}
T_{\mu}^{\,\,\,\rho}&=&\frac{\imath}{4} g^{\rho\nu} \left(\bp
\gamma_\mu
\partial_\nu \psi + \bp \gamma_\nu \partial_\mu \psi -
\partial_\mu \bar \psi \gamma_\nu \psi - \partial_\nu \bp \gamma_\mu
\psi \right)\nonumber\\
& - &\frac{\imath}{4} g^{\rho\nu} \bp \left(\gamma_\mu \G_\nu +
\G_\nu \gamma_\mu + \gamma_\nu \G_\mu + \G_\mu \gamma_\nu\right)\psi
 \,- \delta_{\mu}^{\rho} \left(2 K F_K - F(K)\right). \label{temsp0}
\end{eqnarray}

Finally, exploiting the explicit form of spin connection
(\ref{saclrsbi}) after some manipulations one finds the following
non-trivial components of the energy-momentum tensor of the spinor
field
\begin{subequations}
\label{emtlrsbi}
\begin{eqnarray}
T_0^0 & = & m_{\rm sp} S + F(K), \label{emt00lrsbi}\\
T_1^1 &=& T_2^2 = T_3^3 = F(K) - 2 K F_K, \label{emtiilrsbi}\\
T_3^1 &=&\frac{\imath}{4} \frac{a_3}{a_1} \left(\frac{\dot a_3}{a_3}
- \frac{\dot a_1}{a_1}\right) \bp \bg^3 \bg^1 \bg^0 \psi =
\frac{1}{4} \frac{a_3}{a_1}
\left(\frac{\dot a_3}{a_3} - \frac{\dot a_1}{a_1}\right)\,A^2, \label{emt13}\\
T_3^2 &=&\frac{\imath}{4} \frac{a_3}{a_1} \left(\frac{\dot a_1}{a_1}
- \frac{\dot a_3}{a_3}\right) \bp \bg^2 \bg^3 \bg^0 \psi =
\frac{1}{4} \frac{a_3}{a_1} \left(\frac{\dot a_1}{a_1} - \frac{\dot
a_3}{a_3}\right)\,A^1. \label{emt23}
\end{eqnarray}
\end{subequations}
So the complete set of Einstein equations for a BI metric should be
\begin{subequations}
\label{lrsBIEn}
\begin{eqnarray}
\frac{\ddot a_3}{a_3} +\frac{\ddot a_1}{a_1} + \frac{\dot
a_3}{a_3}\frac{\dot a_1}{a_1}&=& \kappa (F(K) - 2 K F_K),\label{11lrsbin}\\
2\frac{\ddot a_1}{a_1}  +  \frac{\dot
a_1^2}{a_1^2}&=&  \kappa (F(K) - 2 K F_K),\label{33lrsbin}\\
\frac{\dot a_1^2}{a_1^2} + 2 \frac{\dot a_3}{a_3}\frac{\dot
a_1}{a_1}&=& \kappa (m_{\rm sp} S + F(K)), \label{00lrsbin}\\
0 &=& \left(\frac{\dot a_3}{a_3} - \frac{\dot
a_1}{a_1}\right)\,A^2,,
\label{13lrsbin}\\
0 &=& \left(\frac{\dot a_1}{a_1} - \frac{\dot a_3}{a_3}\right)\,A^1.
\label{23lrsbin}
\end{eqnarray}
\end{subequations}

Before solving the Einstein equations let us first write the
equations for the bilinear spinor forms. Recalling that there are 16
bilinear spinor forms, namely, $S = \bar \psi \psi, \quad P = \imath
\bar \psi \gamma^5 \psi, \quad
 v^\mu = \bar \psi \gamma^\mu \psi, \quad  A^\mu = \bar \psi \gamma^5
 \gamma^\mu \psi$, and $Q^{\mu\nu} = \bar \psi \sigma^{\mu\nu} \psi$
 are the scalar, pseudoscalar, vector, pseudovector and
antisymmetric tensor, respectively, for the LRS BI metric one finds
the following system of equations:
\begin{subequations}
\label{invlrsbi}
\begin{eqnarray}
\dot S_0  +  {\cG} A_{0}^{0} &=& 0, \label{S0lrsbi} \\
\dot P_0  -  \Phi A_{0}^{0} &=& 0, \label{P0lrsbi}\\
\dot A_{0}^{0}  +  \Phi P_0 -  {\cG}
S_0 &=& 0, \label{A00lrsbi}\\
\dot A_{0}^{3}  &=& 0, \label{A03lrsbi}\\
\dot v_{0}^{0}  &=& 0,\label{v00lrsbi} \\
\dot v_{0}^{3} +
\Phi Q_{0}^{30} +  {\cG} Q_{0}^{21} &=& 0,\label{v03lrsbi}\\
\dot Q_{0}^{30} -  \Phi v_{0}^{3} &=& 0,\label{Q030lrsbi} \\
\dot Q_{0}^{21} -  {\cG} v_{0}^{3} &=& 0, \label{Q021lrsbi}
\end{eqnarray}
\end{subequations}
where we denote $S_0 = S V,\, P_0 = P V,\, A_0^\mu = A_\mu V,\,
v_0^\mu = v^\mu V,\, Q_0^{\mu \nu} = Q^{\mu \nu} V$ and $\Phi =
m_{\rm sp} + {\cD}$. We also denote ${\cD} = 2 S F_K K_I$ and ${\cG}
= 2 P F_K K_J$, with $F_K = dF/dK$, $K_I = dK/dI$, and $K_J =
dK/dJ.$

Here we also introduce the volume scale
\begin{equation}
V = a_1^2 a_3. \label{VDeflrsbi}
\end{equation}

\section{Solution to the field equations}

From (\ref{S0lrsbi}) - (\ref{Q021lrsbi}) one finds the following
relations:
\begin{subequations}
\label{inv0lrsbi}
\begin{eqnarray}
(S_{0})^{2} + (P_{0})^{2} + (A_{0}^{0})^{2}  &=&
C_1 = {\rm Const}, \label{inv01lrsbi}\\
A_{0}^{3} &=& C_2 = {\rm Const}, \label{inv02lrsbi}\\
v_{0}^{0} &=& C_3 = {\rm Const}, \label{inv03lrsbi}\\
(Q_{0}^{30})^{2} + (Q_{0}^{21})^{2} + (v_{0}^{3})^{2}
 &=& C_4 = {\rm Const}. \label{inv04lrsbi}
\end{eqnarray}
\end{subequations}

Let us now go back to the Einstein equations. The off-diagonal
components of Einstein equations (\ref{13lrsbin}) and
(\ref{23lrsbin}) impose the following restrictions either on the
components of the spinor field or on the metric functions:
\begin{subequations}
\label{reslrsbi}
\begin{eqnarray}
A^2  = 0, \quad A^1 &=& 0,\label{resspinlrsbi} \\
\frac{\dot a_3}{a_3} - \frac{\dot a_1}{a_1} &=& 0.
\label{resmetlrsbi}
\end{eqnarray}
\end{subequations}

The restriction (\ref{resmetlrsbi}) leads to  $a_3 = q_0 a_1$ with
$q_0$ being some constant. In this case the system can be described
by a Friedmann-Robertson-Walker (FRW) model from the very beginning.
Here we do not consider this case, which we will address in a later
work, within the scope of a FRW model.

We consider the case when the restriction is imposed on the
components of the spinor field in detail. Subtraction of
(\ref{33lrsbin}) from (\ref{11lrsbin}) gives

\begin{eqnarray}
\frac{\ddot a_3}{a_3} - \frac{\ddot a_1}{a_1} +  \frac{\dot
a_1}{a_1} \left(\frac{\dot a_3}{a_3} - \frac{\dot a_1}{a_1}\right) =
0, \label{sub31}
\end{eqnarray}
that leads to \cite{saha2001a}
\begin{eqnarray}
a_1 = D_1 V^{1/3} \exp{\left(X_1 \int \frac{dt}{V}\right)}, \quad
a_3 = (1/D_1^2) V^{1/3} \exp{\left(-2X_1 \int \frac{dt}{V}\right)}.
\label{metricflrsbi}
\end{eqnarray}
with $D_1$ and $X_1$ being the integration constants. Thus we see
that the metric functions can be expressed in terms of $V$.

The solutions to spinor field equation (\ref{speq1}) in this case
can be presented as \cite{saha2001a}
\begin{eqnarray}
\psi_{1,2}(t) = \frac{C_{1,2}}{\sqrt{V}} \exp{\left(-i\int {\Phi}
dt\right)}, \quad \psi_{3,4}(t) = \frac{C_{3,4}}{\sqrt{V}}
\exp{\left(i\int  {\Phi} dt\right)}, \label{psinl}
\end{eqnarray}

with $C_1,\,C_2,\,C_3$, and $C_4$ being the integration constants
and related to $V_0$ as
$$C_1^* C_1 + C_2^* C_2 - C_3^* C_3 - C_4^* C_4 = V_0.$$ Here we
assumed that $K = I$, i.e., $F = F(I)$. The reason for this choice
is discussed later.

Thus we see that the metric functions, the components of spinor
field as well as the invariants constructed from metric functions
and spinor fields are some inverse functions of $V$ of some degree.
Hence any space-time point where $V = 0$ is a singular point. So it
is important to study the behavior of $V$, which we do in the next
section.

\section{Results and discussion}

In this section we discuss the results obtained in the previous
section. In doing so, we pay special attention to the volume scale,
$V$.

Let us first see whether the model becomes asymptotically isotropic.
It can be shown that for an expanding Universe, when $V \to \infty$
as $t \to \infty$, the isotropization process of the Universe takes
place. To prove that we exploit the isotropization condition
proposed in \cite{Bronnikov}
\begin{equation}
\frac{a_i}{a}\Bigl|_{t \to \infty} \to {\rm const.} \label{isocon}
\end{equation}
Then by rescaling some of the coordinates, we can make $a_i/a \to
1$, and the metric will become manifestly isotropic at large $t$.

Taking into account that $a = V^{1/3}$ from (\ref{metricflrsbi}) we
find
\begin{eqnarray}
\frac{a_1}{a} = D_1 \exp{\left(X_1 \int \frac{dt}{V}\right)} \to
D_1, \quad \frac{a_3}{a} = (1/D_1^2) \exp{\left(-2X_1 \int
\frac{dt}{V}\right)} \to 1/D_1^2,\label{isocon1}
\end{eqnarray}
as $V \to \infty$. Recall that the isotropic FRW model has the same
scale factor in all three directions (i.e., $a_1(t) = a_2(t) =
a_3(t) = a(t)$). So for the LRS BI universe to evolve into a FRW one
we should have $D_1 = 1$. Moreover, the isotropic nature of the
present Universe leads to the fact that  $|X_1| << 1$, so that $
\int [V (t)]^{-1}dt \to 0$ for $t < \infty$ (for $V (t) = t^n$ with
$n > 1$ the integral tends to zero as $t \to \infty$).

Our next step will be to define $V$. Combining the diagonal Einstein
equations (\ref{11lrsbin}) -- (\ref{00lrsbin}) in a certain way for
$V$ we find  \cite{saha2001a}
\begin{eqnarray}
\ddot V = \frac{3 \kappa}{2} \left(m_{\rm sp}\,S +  2 ( F(K) -  K
F_K)\right) V. \label{detvlrsbi}
\end{eqnarray}

Now in order to solve (\ref{detvlrsbi}) we have to know the relation
between the spinor and the gravitational fields. Using the equations
(\ref{S0lrsbi}) and (\ref{P0lrsbi}) it can be show that

\begin{equation}
K = \frac{V_0^2}{V^2}. \label{KV}
\end{equation}

Relation (\ref{KV}) holds only for massless spinor field if $K$
takes one of the expressions $\{J,\,I + J,\,I - J\}$, while for $K =
I$ it holds both for massless and massive spinor fields. In the case
of $K = I + J$ one can write $S = \sin{(V_0/V)}$ and $P =
\cos{(V_0/V)}$, whereas for $K = I - J$ one can write $S =
\cosh{(V_0/V)}$ and $P = \sinh{(V_0/V)}$. In what follows, we will
consider the case for $K = I$, as in this case further setting
spinor mass $m_{\rm sp} = 0$ we can revive the results for other
cases. Assuming
\begin{equation}
F = \sum_{k} \lambda_k I^{n_k} =  \sum_{k} \lambda_k S^{2 n_k}
\label{nonlinearityIII}
\end{equation}
on account of $S = V_0/V$ we find

\begin{eqnarray}
\ddot V = \frac{3 \kappa}{2} \left[m_{\rm sp}\,V_0 + 2 \sum_{k}
\lambda_k( 1 - n_k) V_0^{2n_k} V^{1 - 2n_k}\right],
\label{detvlrsbinew}
\end{eqnarray}
with the solution in quadrature
\begin{eqnarray}
\int \frac{dV}{\sqrt{3 \kappa \left[m_{\rm sp} V_0 V + \sum_{k}
\lambda_k V_0^{2n_k} V^{2(1 - n_k)}\right] + {\bar C} }} =  t + t_0,
\label{quadlrsbi}
\end{eqnarray}
with ${\bar C} $ and $t_0$ being some arbitrary constants.

Thus we see that the metric functions, the components of the spinor
field, as well as the invariants constructed from metric functions
and spinor fields are some inverse functions of $V$ of some degree.
Hence any space-time point where $V = 0$ is a singular point. So we
consider that the initial value of $V (0)$ is small but non-zero. As
a result for the nonlinear term to prevail in \eqref{detvlrsbinew}
we should have $ n_k = n_1: 1 - 2 n_1 < 0$ (i.e., $n_1
> 1/2$) whereas for an expanding Universe when $V \to \infty$ as $t
\to \infty $ one should have $ n_k = n_2: 1 - 2 n_2 > 0$  (i.e.,
$n_2 < 1/2$). As is seen from \eqref{detvlrsbinew}, $n_k = n_0: n_0
= 1/2$ leads to a term that can be added to the mass term.

In this case we obtain

\begin{eqnarray}
\ddot V &=& \Phi_1(V), \label{detvlrsbinew1}\\
\Phi_1(V) &=& \frac{3 \kappa}{2} \left[\left(m_{\rm sp} +
\lambda_0\right)\,V_0 + 2\lambda_1( 1 - n_1)
 V_0^{2n_1} V^{1 - 2n_1} + 2\lambda_2( 1 - n_2)
 V_0^{2n_2} V^{1 - 2n_2}\right].\nonumber
\end{eqnarray}
Equation \eqref{detvlrsbinew1} allows the first integral
\begin{eqnarray}
\dot V &=& \Phi_2(V), \label{1stintlrsbi}\\
\Phi_2(V) &=& \sqrt{3 \kappa \left[\left(m_{\rm sp} +
\lambda_0\right) V_0 V +\lambda_1
 V_0^{2n_1} V^{2(1 - n_1)} + \lambda_2
 V_0^{2n_2} V^{2(1 - n_2)} +
{\bar C}\right]}.
\end{eqnarray}
The solution to
 \eqref{detvlrsbinew1} can be written in quadrature
as follows:
\begin{eqnarray}
\int \frac{dV}{\Phi_2(V)} = t + t_0. \label{quadlrsbi1}
\end{eqnarray}

To solve \eqref{detvlrsbinew1} we should choose the problem
parameters $V_0$, $m_{\rm sp}$, $\kappa$, $\bar C$, $\lambda_k$, as
well as the initial value of $V(0)$ in such a way that does not lead
to
$$ \left(m_{\rm sp} +
\lambda_0\right) V_0 V +\lambda_1
 V_0^{2n_1} V^{2(1 - n_1)} + \lambda_2
 V_0^{2n_2} V^{2(1 - n_2)} +
{\bar C} < 0.$$

For simplicity let us set $V_0 = 1$,\, $m_{\rm sp} = 1$,\, $C_0 =
10$, and $\kappa = 1$. In line with our discussion earlier we
consider $n_0 = 1/2$, $n_1 = 2$, and $n_2 = 0$. In our case we set
$V(0) = 0.5$. We set $\lambda_0 = 1$, whereas $\lambda_1 = \pm 1$
and $\lambda_2 = \pm 1$ were taken in different combinations. It was
found that depending on the sign of $\lambda_2$ the model gives
principally different types of solutions, namely, in the case of
positive $\lambda_2$ we have an accelerated mode of expansion of the
Universe, while for negative $\lambda_2$ we have oscillatory
solution.

Defining the deceleration parameter
\begin{equation}
q = - \frac{V \ddot V}{{\dot V}^2} = -\frac{V \Phi_1 (V)}{\Phi_2^2
(V)}, \label{decellrsbi}
\end{equation}
from \eqref{detvlrsbinew1} and \eqref{1stintlrsbi} we have
\begin{equation}
q = - \frac{ \frac{3 \kappa}{2} \left[\left(m_{\rm sp} +
\lambda_0\right)\,V_0 V + 2\lambda_1( 1 - n_1)
 V_0^{2n_1} V^{2(1 - n_1)} + 2\lambda_2( 1 - n_2)
 V_0^{2n_2} V^{2(1 - n_2)}\right]}{3 \kappa \left[\left(m_{\rm sp} +
\lambda_0\right) V_0 V +\lambda_1
 V_0^{2n_1} V^{2(1 - n_1)} + \lambda_2
 V_0^{2n_2} V^{2(1 - n_2)} +
{\bar C}\right]}. \label{decellrsbinew}
\end{equation}

Taking into account that for an expanding Universe at large $t$ the
term $V^{1 - 2n_2}$ prevails, for deceleration parameter we find
\begin{equation}
\lim_{V \to \infty}q  \longrightarrow - (1 - n_2) < 0, \quad {\rm
since} \quad n_2 < 1/2. \label{accele}
\end{equation}
Thus we see that spinor field nonlinearity generates late time
acceleration of the Universe.

\begin{figure}[ht]
\centering
\includegraphics[height=70mm]{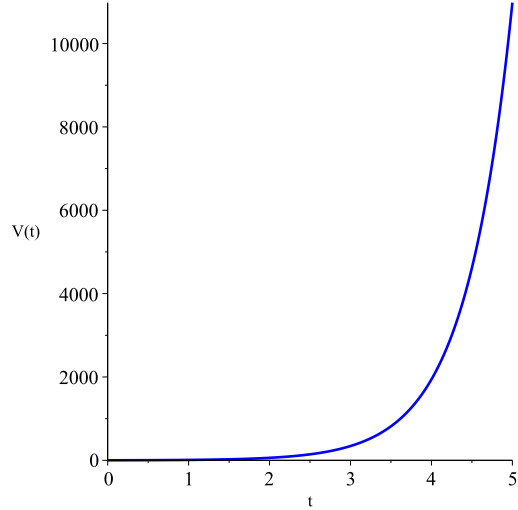} \\
\caption{Evolution of the Universe for a positive  $\lambda_2$}
\label{Vlrsbipos}
\end{figure}

\begin{figure}[ht]
\centering
\includegraphics[height=70mm]{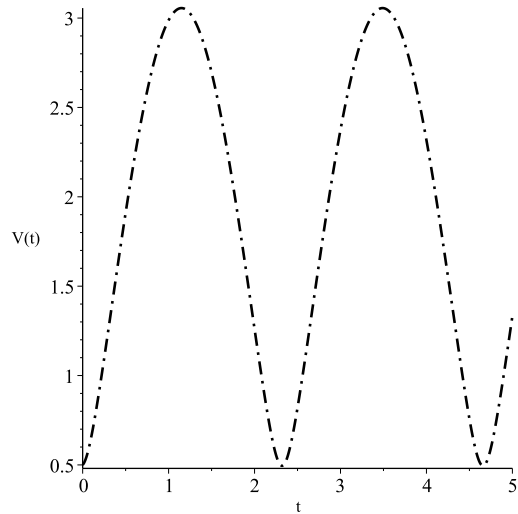} \\
\caption{Evolution of the Universe for a negative  $\lambda_2$}
\label{Vlrsbineg}
\end{figure}

\begin{figure}[ht]
\centering
\includegraphics[height=70mm]{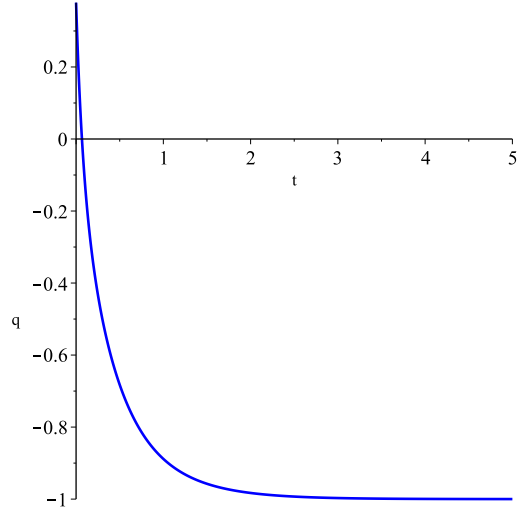} \\
\caption{Plot of deceleration parameter $q$ for a positive
$\lambda_2$ } \label{qlrsbipos}
\end{figure}

In Fig. \ref{Vlrsbipos} and Fig. \ref{Vlrsbineg} we plotted the
evolution of volume scale $V$ for a positive and negative
self-coupling constant $\lambda_2$, respectively. As one sees from
Fig. \ref{Vlrsbipos}, a positive $\lambda_2$ gives rise to an
accelerated mode of expansion, whereas Fig. \ref{Vlrsbineg} with
negative $\lambda_2$ shows the oscillatory mode of expansion. In
Fig. \ref{qlrsbipos} the deceleration parameter $q$ is illustrated
for a positive $\lambda_2$.

Finally we study what happens to shear and anisotropic parameters in
this case. In doing so, let us first rewrite the corresponding
quantities. The expansion $\vartheta$ for LRS BI metric reads
\begin{equation}
\vartheta = 2\frac{\dot a_1}{a_1} + \frac{\dot a_3}{a_3} =
\frac{\dot V}{V}, \label{explrsbi}
\end{equation}
whereas, from
\begin{subequations}
\label{shearcompslrsbi}
\begin{eqnarray}
\sigma_{1}^{1} &=& \sigma_{2}^{2} =  \frac{\dot a_1}{a_1} - \frac{1}{3} \vartheta
= \frac{1}{3} \left(\frac{\dot a_1}{a_1} - \frac{\dot a_3}{a_3}\right), \label{sh11lrsbi}\\
\sigma_{3}^{3} &=&  \frac{\dot a_3}{a_3} - \frac{1}{3} \vartheta = -
\frac{2}{3} \left(\frac{\dot a_1}{a_1} - \frac{\dot a_3}{a_3}\right)
\label{sh33lrsbi}
\end{eqnarray}
\end{subequations}
one finds the expression for shear
\begin{equation}
\sigma^ 2 = \frac{1}{2}\left[\sum_{i=1}^3 \left(\frac{\dot
a_i}{a_i}\right)^2 - \frac{1}{3}\vartheta^2\right] =  \frac{1}{3}
\left(\frac{\dot a_1}{a_1} - \frac{\dot a_3}{a_3}\right)^2.
\label{sheargenlrsbi}
\end{equation}
The anisotropic parameter in this case has the form
\begin{equation}
A_m = \frac{1}{3}\sum_{i=1}^{3}\left(\frac{H_i}{H} - 1\right)^2 =
\frac{1}{3 H^2}\left[2\left(\frac{\dot a_1}{a_1}\right)^2 +
\left(\frac{\dot a_3}{a_3}\right)^2\right]
 - 1, \label{anisoparam}
\end{equation}
where $H = \frac{1}{3}\left(2\frac{\dot a_1}{a_1} + \frac{\dot
a_3}{a_3}\right) = \frac{1}{3}\frac{\dot V}{V}$. Further from
\eqref{metricflrsbi} we find that
\begin{equation}
\frac{\dot a_1}{a_1} = \frac{1}{3} \frac{\dot V}{V} + \frac{X_1}{V},
\quad \frac{\dot a_3}{a_3} = \frac{1}{3} \frac{\dot V}{V} -2
\frac{X_1}{V}. \label{Hubbledir}
\end{equation}
Now inserting \eqref{Hubbledir} into \eqref{shearcompslrsbi} --
\eqref{anisoparam} we finally find
\begin{equation}
\sigma_{1}^{1} = \sigma_{2}^{2} = \frac{X_1}{V}, \quad
\sigma_{3}^{3} = -2 \frac{X_1}{V}, \label{shearcom01}
\end{equation}
\begin{equation}
\sigma^2 = 3 \frac{X_1^2}{V^2}, \label{shear01}
\end{equation}
\begin{equation}
A_m = 18 \frac{X_1^2}{\dot V^2}. \label{anisoparam01}
\end{equation}
As was shown in \eqref{detvlrsbinew1} and \eqref{1stintlrsbi} in the
case of a positive $\lambda_2$ that generates late time
acceleration, both $V$ and $\dot V$ become large with the expansion
of the Universe leading to $\sigma_{i}^{i} \to 0$,   $\sigma^2 \to
0$ and $A_m \to 0$. This corresponds to our earlier conclusion
regarding isotropization. As far as negative $\lambda_2$ is
concerned, in this case the model gives rise to an oscillatory mode
of expansion. In this case we have both local minima and maxima. The
maximum (minimum) value of volume scale $V$ depends on the
parameters and the initial condition and  may be as large as
possible. Hence in case of a negative $\lambda_2$ though it is
possible to attain a solution such that $\sigma_{i}^{i}|_{V = V_{\rm
max}} \to 0$ and $\sigma^2|_{V = V_{\rm max}} \to 0$, but at the
same time we have ${A_m}|_{V = V_{\rm max}} \to \infty$ because at
any space-time point where $V = V_{\rm max (min)}$ we have $\dot
V|_{V = V_{\rm max (min)}} = 0$. This means that at any space-time
point where evolution changes its direction (expansion to
contraction and vice versa) the Universe becomes highly anisotropic.

\section{Conclusion}

Within the scope of the LRS BI cosmological model we studied the
role of the spinor field in the evolution of the Universe. The
reason for considering the LRS BI model lies in the fact that in the
case of a full BI model the non-diagonal components of the
energy-momentum tensor of the spinor fields imposes severe
restrictions on the components of the spinor field, resulting in
vanishing scalar $S = \bp \psi$ and pseudoscalar $P = \imath \bp
\gamma^5 \psi$ \cite{sahaIJTP2014,sahaAPSS2015}. As a result both
the mass term and nonlinear term in the Lagrangian disappear. But,
as was shown here, in the case of an LRS BI cosmological model,
neither the mass term nor the nonlinear term vanish. Moreover,
unlike the Bianchi type-VI model the present model leads to
asymptotic isotropization. It is was also found that, depending on
the sign of the self-coupling constant, the model allows either the
accelerated mode of expansion or the oscillatory mode of evolution.

\vskip 0.1 cm

\noindent {\bf Acknowledgments}\\
This work is supported in part by a joint Romanian-LIT, JINR, Dubna
Research Project, theme no. 05-6-1119-2014/2016.

\end{document}